\documentclass
[aps,prl,twocolumn,showpacs,preprintnumbers,amsmath,amssymb,superscriptaddress]{revtex4}

\usepackage{graphicx}
\usepackage{dcolumn}
\usepackage{amsmath,amssymb}

\newcommand{\bm}[1]{\mbox{\boldmath{$#1$}}}

\makeatletter
\def\btt#1{\texttt{\@backslashchar#1}}%
\DeclareRobustCommand\bblash{\btt{\@backslashchar}}
\makeatother

\begin{document}

\title[Short Title]{
Degeneracy and Consistency Condition of Berry phases:\\
Gap Closing under the Twist 
}
\author{T. Hirano}
\email{hirano@pothos.t.u-tokyo.ac.jp}
\affiliation{%
Department of Applied Physics, The University of Tokyo, 7-3-1 Hongo,
Bunkyo-ku, Tokyo 113-8656, Japan }%
\author{H. Katsura}%
\email{katsura@appi.t.u-tokyo.ac.jp}
\affiliation{%
Department of Applied Physics, The University of Tokyo, 7-3-1 Hongo,
Bunkyo-ku, Tokyo 113-8656, Japan }%
\author{Y. Hatsugai}%
\thanks{Corresponding author}
\email{hatsugai@sakura.cc.tsukuba.ac.jp}
\affiliation{%
Department of Applied Physics, The University of Tokyo, 7-3-1 Hongo,
Bunkyo-ku, Tokyo 113-8656, Japan }%
\affiliation{%
Institute of Physics, 
University of Tsukuba, 1-1-1 Tennodai, Tsukuba, Ibaraki 305-8571, Japan}
\date{\today}
\begin{abstract}
We have discussed a consistency condition of 
 Berry phases defined by a 
local gauge twist and spatial symmetries of the 
many body system.
It imposes a non trivial gap closing condition under the gauge twist in both finite- and infinite-size systems. 
It also implies a necessary condition for the
gapped and unique ground state.
As for the simplest case, 
it predicts an inevitable 
gap closing in the Heisenberg chain of half integer spins. 
Its relation to the Lieb-Schultz-Mattis theorem is 
discussed based on the symmetries of the twisted Hamiltonian. 
The discussion is also extended to the (approximately) 
degenerated multiplet and fermion cases.
It restricts the number of the states in the low energy cluster
of the spectrum by the filling of the fermions.
Constraints by the reflection symmetry are also discussed.
%
\end{abstract}
\maketitle
For the last decade,
quantum criticality with gapless excitations have been focused 
in its relation to a quantum phase transition.
Generically speaking, the gapless phase 
is only realized by a fine 
tuning of the parameters of the quantum hamiltonian.
Some machinery to protect the gap closing of the 
quantum mechanical system is necessary.
Symmetries as a spatial translation can be one of 
the reasons which plays an important role in the Fermi liquids.
Today we have other several machinery for the gap closing
in a generic situation.
One is the Nambu-Goldstone mechanism.
When a continuous symmetry is broken spontaneously, 
there can be 
a gapless excitation by slowly varying its local order parameter
which is responsible for the symmetry breaking. 
It is closely related to 
the Lieb-Schultz-Mattis(LSM)
argument\cite{LSM,Affleck-Lieb,AffleckLSM,OshikawaLSM,YamanakaLSMluttinger,HastingsLSM,MisguichLSM}
 to prove  existence of the gapless excitations in a
half-odd-integer spin chains.
This LSM argument is justified in  restricted situations.
It allows  existence of  a gapped phase
which is known as the Haldane phase in the integer spin chains. 
This is a typical quantum liquid where strong quantum fluctuations
prevent from forming an ordered phase.
The other mechanism to protect the gapless excitations is 
an appearance of the edge states such as the one in quantum Hall states
and Haldane
phases\cite{LaughlinEdge,HalperinEdge,HatsugaiER,HatsugaiERC,
HaldaneGap,KennedyEdge,Hagiwara,NENP1,NENP2}.
There is a fundamental class of physical phases with intrinsic 
energy gap as topological insulators. They are gapped quantum liquids
in a bulk without any fundamental symmetry breaking.
However the system has a geometrical (topological) perturbation 
such as the existence of the boundaries,  low energy modes (quasi-particle)
 as the generic edge states appear and the phase becomes gapless
when the boundaries are infinitely separated. 

The existence of the energy gap is also closely related to the 
degeneracy of the ground state.
When the discrete symmetry is broken such as for the Heisenberg spins 
with ising anisotropy
or charge ordered states, the symmetry broken states split into
a low energy cluster
with small energy splitting 
in a finite periodic system.
The exact degeneracy is only realized in the infinite size limit.
For the topological insulators without any symmetry breaking, 
there can be additional degeneracy as the topological degeneracy
which characterizes the non trivial quantum liquids\cite{TOrder,Wu}.

As for the topological insulators, 
quantum order parameters by the many body Berry phases, associated with
the local gauge twists,
have been  proposed and its validity is justified for
 several concrete models
\cite{HatsugaiOrder123,Maruyama,VBSBerry}.
Using this Berry phase and its symmetry property,
we prove that there is an inevitable degeneracy 
during the twist as for some classes of Hamiltonian
which is also related to the LSM theorem. 
Our discussions are valid not only in infinite-size systems but also in finite-size systems, which can be applicable for various correlated electron systems.

Let us define a local order parameter $\gamma_{ij}$ at
a link $(ij)$ by the Berry phase\cite{Berryphase}
$
i\gamma_{ij}=\int_{0}^{2\pi}d\phi A_{ij}(\phi)
$,  
where $A_{ij}(\phi)$ is a  Berry connection obtained from a
gauge fixed 
(single valued) normalized ground state $|\psi_{ij}\rangle$ 
of a Hamiltonian $H_{ij}(\phi)$ as 
$A_{ij}(\phi)=\langle\psi_{ij}|\partial_{\phi}|\psi_{ij}\rangle$
\cite{HatsugaiOrder123},
where $\phi$ is a parameter of
the local $U(1)$ twist on the link.
It is gauge dependent but is well defined up to modulo $2\pi$. 
Further it is  quantized if the ground state is 
invariant under some anti-unitary operation.
As for the Heisenberg model with generic connectivity,
 the Hamiltonian with the local twist is given as
$H_{ij}(\phi)
=
J_{ij}\left((e^{i\phi}S_{i}^+S_{j}^-+e^{-i\phi}S_{i}^-S_{j}^+)/2+S_i^zS_{j}^z\right)
+\sum_{(kl)\neq (ij)}J_{kl} \bm{S}_k\cdot\bm{S}_{l}
$.
We should note that 
the Berry phase is only well defined
unless the energy gap
between the ground state and the first excited
states does not vanish during the twist.

Since the twist on the the link can be modified 
by the local gauge transformation, it
gives some constraints for the Berry phase distribution. 
For simplicity, let us first discuss the 
one dimensional nearest neighbor antiferromagnetic
Heisenberg model of generic spins on a finite lattice.
We discuss within
 the subspace of fixed $S_{tot}^z=\sum S_{j}^z$ since it commutes with $H_{i,j}(\phi)$. 
Performing a local gauge transformation at the site $j$
as 
\begin{alignat*}{1} 
U_{j}(\phi) =& e^{i(S-S_{j}^z)\phi},\quad \phi\in[0,2\pi]
\end{alignat*}   
which is single valued in the parameter space, 
one has an important  relation between the two different hamiltonians
(which are gauge equivalent),
$H_{j,j+1}$ and $H_{j-1,j}$ 
as
$H_{j,j+1}= U_{j}^{\dagger} H_{j-1,j}U_{j}$.
Correspondingly the states are mapped into each other as 
$|\psi_{j,j+1}\rangle=U_{j}^{\dagger}|\psi_{j-1,j}\rangle$,
where $|\psi_{j-1,j}\rangle$ and $|\psi_{j,j+1}\rangle$ 
are two different ground states of $H_{j-1,j}$ and $H_{j,j+1}$,
respectively.
Note that,
if the state $|\psi_{j-1,j}\rangle$ is gauge fixed as a single valued in the
parameter space,
 it is also true for the state $|\psi_{j,j+1}\rangle$.
Assuming that 
{\it the ground state is unique and gapped during the twisting},
this simple relation gives a constraint for the quantum local order parameters
$\gamma_{ij}$'s
\begin{eqnarray*}
\gamma_{j,j+1}
&=&
-i \int \langle\psi_{j-1,j}|U_{j}\partial_{\phi} 
(U_{j}^{\dagger}|\psi_{j-1,j}\rangle) \,d\phi,
\\
&=&  \gamma_{j-1,j} +
\int (S-\langle \psi|S_{j}^z|\psi\rangle) \, d\phi.
\end{eqnarray*}
Since the expectation value of the hermitian operator $S_j^z$ is 
independent of the gauge transformation, 
we have dropped the label of the wave function
 to specify the position of the twist. 
The time reversal invariance of the state $|\psi\rangle$ implies 
$\langle\psi|S_i^z|\psi\rangle=
-\left(\langle\psi|S_i^z|\psi\rangle\right)^*=0$.
Then, one obtains 
\begin{eqnarray}
\gamma_{j-1,j}&=&2\pi S + \gamma_{j,j+1}.
\nonumber
\end{eqnarray}
Since the present 1D Heisenberg model has a translational symmetry, 
the Berry phases as the quantum order parameters should also 
respect this  translational symmetry, that is,
$\gamma _{ij}$ should be independent of the link $(ij)$ in mod $2\pi$
unless they are well defined.
However this is impossible for the half integer spins due to
the above constraint $\gamma_{j-1,j}\equiv\gamma _{j,j+1}+\pi,
\  ({\rm mod} 2\pi)$.
It implies that 
{\em
there is a gap closing under the local twist,
as for the antiferromagnetic Heisenberg chains 
with half-integer spins on a finite lattice.
}
Although our results can be applicable for arbitrary half integer spins,
similar conclusions for the $S=1/2$ case are 
also obtained from different analysis
\cite{Kolb,KorepinBerry,MisguichLSM}.
As for the integer spin Heisenberg model, the above constraint does not
forbid uniform distribution of the Berry phase $\gamma _{ij}$. Actually
the uniform $\pi$ Berry phase is realized in a $S=1$
case\cite{HatsugaiOrder123,HatsugaiOrderP,VBSBerry}.

The idea here is extended to the general 
cases without any difficulty.
Let us consider sets of some generic twists for several links labeled by ``In''.
Assuming the uniqueness of the ground state even under the twisting,
one can define the Berry phase $\gamma _{\rm In}$. 
Next let us perform a set of the local 
gauge transformation within the area ${\cal A}$.
Then  we have
a new set of twists labeled by ``Out''.
Under this generic setup, one has a constraint
between the two corresponding Berry phases
 $\gamma _{\rm In}$ and $\gamma _{\rm Out}$
as
\begin{eqnarray*}
\gamma_{\rm In}&=&\gamma_{\rm Out}
+2\pi\sum_{i\in{\cal A}}\left(S_i-\tilde{m}_i\right),\ 
\tilde{m}_i=\frac{1}{2\pi}\int\langle\psi|S_i^z|\psi\rangle d\phi.
\end{eqnarray*}

\begin{figure}[!tb]
 \includegraphics[width=7.5cm]{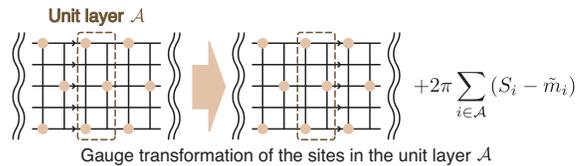}
\caption{The definition of the unit layer, and the gauge transformation
 of the Berry phase.
The arrows on links denotes the link twist.}
\label{fig:UnitLayer}
\end{figure}

\paragraph{Translational symmetry}
For a spin ladder as an example of translational invariant
case, 
we take ${\cal A}$ as a unit layer and 
 define Berry phase by twisting the links on the left boundary of the
area $\cal A$ simultaneously (see FIG.\ref{fig:UnitLayer}).
Due to the translational symmetry, 
we have a constraint for the Berry phases as $\gamma_{\rm In}
\equiv \gamma_{\rm Out}$ (mod $2\pi$)
assuming that the ground state is gapped under the twisting.
Since the total $S^z$ is conserved,
we have $\sum_{i\in{\cal A}}\tilde m_i=(1/2\pi )
\int d\phi \langle\psi|\sum_{i}S^z_i|\psi \rangle/N
=\langle\psi|\sum_{i}S^z_i|\psi \rangle/N=|{\cal A}|m$, 
where $|{\cal A}|$ is a number of sites in the unit layer and
$m$ is the average magnetization
($N$ is a number of the unit layers).
Note that the translational invariance of the magnetization at
arbitrary $\phi$ is guaranteed
by the fact that $S^z_{total}$ commutes with a unitary transformation 
which spreads the flux in a transitionally invariant way. 
Then we have a necessary condition for the unique and gapped ground state as
\begin{eqnarray*}
 \sum_{i\in{\cal A}}{(S_i-m)}& \equiv& \nu\in\mathbb{Z}
\end{eqnarray*}
It is a condition  
to have a magnetization plateau\cite{OshikawaLSMplateau}.

By considering a non-Abelian Berry phase\cite{HatsugaiOrder123}, 
this argument can be extended to the case with (approximate) 
degeneracy.
Now let us assume the low energy spectrum near the ground states
forms a multiplet $\Psi=(|\psi_1\rangle,\cdots,|\psi_M\rangle)$ 
with $M$ eigenstates,
 $|\psi_i\rangle$ ($i= 1,\cdots,M$).
Here, we assume that they are in the subspace of the same $S^z_{tot}$
and the energy gap above the multiplet is stable.
Then  as for the Berry phase
 $\gamma= \int d\phi\,{\rm Tr \,} \Psi ^\dagger \partial _\phi \Psi$,
we have a relation
$\gamma_{\rm In}
=\gamma_{\rm Out}+2\pi M\nu
$ based on the same assumption.
Therefore as for the translational invariant system, 
we have a consistency requirement
$2\pi M\nu\equiv 0$ (mod $2\pi$) assuming the gap even under  the twist.
Then { \em as for $\nu=\sum_{i\in{\cal A}}(S_i-m)=\frac{p}{q}$ with
 mutually co-prime $p$ and $q$ case, $M$ has to be a multiple of $q$, 
that is, the low energy spectrum has to form a cluster of 
$q\ell$ states ($\ell$: integer).}
This situation naturally 
occurs with discrete symmetry breaking \cite{HatsugaiGap-opening} or
 topological degeneracy.

As for the $S=1/2$ Heisenberg ladder with $n$ legs,
 the  discussion here
predicts a gap closing under the twist when $n$ is odd\cite{DagottoLad,Aringa}.
It allows to have an energy gap
above the low energy multiplet composed of two states
even with the $S=1/2$ system, 
which is realized for the spin tube\cite{SpinTube}.
The present argument also gives a consistent description for ferrimagnets
discussed by the effective field theory and the LSM
 argument\cite{Fukui-Kawakami}.
Consider a  Heisenberg spin chain with different spin quantum numbers as 
$S_j=S_1$ for $j=1\mbox{ (mod $M$)}$ and $S_j=S_2$ for others.
Taking the unit layer to include these $M$ spins,
the gauge transformation yields the relation of the Berry phase as
$\gamma_1=\gamma_2+2\pi(S_1+(M-1)S_2)$.
The system must have the gapless excitations or the ground state
degeneracy if $(S_1+(M-1)S_2)\not\in\mathbb{Z}$, and
the system can have a unique and gapped ground state if
$(S_1+(M-1)S_2)\in\mathbb{Z}$.

We shall now discuss the connection to the LSM theorem. 
Our argument suggests that there is at least one level-crossing point
during the local twist if Berry phases cannot be arranged in a
compatible way with the translational symmetry. Indeed, one can
rigorously show the degeneracy of the ground state at $\phi=\pi$. For simplicity, we consider the first example, namely the half-integer Heisenberg spin chain with length $L$ in a zero magnetic field. 
We introduce the following two symmetry operations. One is $U_j(\phi) T$, where $T$ is the operation for the one-step translation along the chain. The Hamiltonian $H_{j-1,j}$ is invariant under this operation: $(U_j(\phi)T)^\dagger H_{j-1,j}(\phi)(U_j(\phi)T)=H_{j-1,j}(\phi)$.  
The other is the spin flip operation $F$ defined by $FS^z_jF=-S^z_j$ and $FS^{\pm}_jF=S^{\mp}_j$ for any $j$. The Hamiltonian has this symmetry if and only if $\phi=0$ or $\pi$, {\it i.e.}, $FH_{j-1,j}(\phi=0/\pi)F=H_{j-1,j}$. 
At $\phi=\pi$, there is a hidden algebraic relation: $\{U_j(\pi)T,F\}=0$ where $\{,\}$ denotes the anticommutator. This can be shown by the fact that $FU_j(\pi)F=U_j(\pi)e^{2\pi i S^z_j}=-U_j(\pi)$ since we consider the half-odd-integer spin chains. From the anticommutation relation, there exist at
least two ground states labeled by the quantum number associated to $F$ at $\phi=\pi$. 
We call two of them $|G(\pi,+1)\rangle$ and $|G(\pi,-1)\rangle \equiv (U_j(\pi)T)|G(\pi,+1)\rangle$, where $F|G(\pi,\eta)\rangle =\eta|G(\pi,\eta)\rangle$. 
This degeneracy is not restricted to the ground state. From our argument, it is obvious that every energy level is at least doubly degenerate at $\phi=\pi$ and can be distinguished by the eigenvalue of $F$.
An extension to the above argument to other systems with
translational symmetry can be done in a straightforward way by replacing
$U_j(\phi)T$ with $U_{\cal A}(\phi)T_{\cal A}$, where $U_{\cal
A}(\phi)=\prod_{j\in {\cal A}}U_j(\phi)$ and $T_{\cal A}$ is the
translation of the unit layer. 

To discuss the relation between this degeneracy and the LSM theorem, it is useful to introduce the translationally invariant Hamiltonian with the twist $\phi$ as ${\tilde H}(\phi)\equiv U^{\dagger}(\phi)H_{L,1}(\phi)U(\phi)$, where $U(\phi)=\prod^L_{j=1}U_j(-j \phi/L)$. 
The level-crossing at $\phi=\pi$ suggests that one of the excited state of ${\tilde H}(0)=H_{L,1}(0)$ is smoothly connected to the ground state of ${\tilde H(2\pi)}$. 
Since $H_{L,1}(2\pi)=H_{L,1}(0)$, the ground state of ${\tilde H}(2\pi)$ is given by $U^{\dagger}(2\pi)|G(0)\rangle$, where $|G(0)\rangle$ is the ground state of ${\tilde H}(0)$. Finally, along the same lines as the LSM argument, one can show that $U^{\dagger}(2\pi)|G(0)\rangle$ is orthogonal to $|G(0)\rangle$ and the energy difference between them is ${\cal O}(1/L)$ using the translational and $F$ symmetry. 

Another comment is that,
in the case of one dimensional spin chains with open boundary condition,
 the local gauge twists are always gauged away.
Using the gauge transformation of the string type, 
$\prod_{i=1}^{j}U_{i}(\phi)=e^{i\sum_{i=1}^j(S-S_{i}^{z})\phi}$,
we obtain the Berry phase as $\gamma_{j,j+1}=2\sum_{i=1}^{j}S_{i}\pi$
assuming the energy gap.
It is consistent with the generic VBS state 
\cite{VBSBerry}.
As for the $S=1$ spin chain 
with open boundaries,
the Haldane gap corresponds to the energy gap 
above the Kennedy triplet\cite{KennedyEdge}.
Then the Berry phase
 of the low energy cluster below the 
Haldane gap, which includes contribution of the edge states,
gives vanishing Berry phase.
It should be distinguished 
from the translationally invariant case 
without edge \cite{HatsugaiOrder123,HatsugaiOrderP}.

 Further this  present argument is also applicable for systems with
charge degrees of freedom. 
Let us consider a fermion model with conserved particle number
 $n_{tot}=\sum_i n_i$ 
such as  spinless fermions with interaction 
 $H=\sum_{\langle ij\rangle}(t c_i ^\dagger c_{j}+ h.c. + V_{ij} n_i n_j)$,
where $n_i=c_i ^\dagger c_i$ and
 $c_i$ is a fermion annihilation operator at site $i$.
The $U(1)$ gauge twist against the charge degree of freedom
is introduced by replacing  a hopping at the special link $(ij)$ as 
$e^{i\phi}c^{\dagger}_ic_j$ and the
 $U(1)$ local gauge transformation is given by
\begin{alignat*}{1}
U_j(\phi) =& e^{i n_j\phi}.
\end{alignat*} 
Then the transformation property of the 
Berry phase 
under the gauge transformation
leads the relation
\begin{alignat*}{1} 
\gamma_{\rm In} =& \gamma_{\rm Out}
-2\pi\sum_{i\in{\cal A}}\tilde{n}_{i},
\end{alignat*} 
where $\tilde{n}_i=\frac{1}{2\pi}\int \langle\phi|n_i|\phi\rangle d\phi$ 
as the case of spins and $\cal A$ is a unit layer.
Following the same argument as the spin case,
the  translational symmetry 
gives a requirement for the gap under the twist
as 
\begin{alignat*}{1} 
\rho &  \in\mathbb{Z},
\ \
\rho = 
\frac {1}{N} \langle \sum_{j} n_j \rangle
\end{alignat*} 
where $\rho$ is an average 
particle number per unit layer and $N$ is a number of the unit layers.
It has also a non-Abelian extension for the degenerate multiplet
which is just a repetition of the spin case. 
It requires that,
when the filling is $\rho=p/q$ with mutual co-prime $p$ and $q$,
there exists a multiplet of  $M$ states in the low energy spectrum to
form a cluster,
which is separated from the else under the
twist as
$M 
=q\ell,\ \ell=1,2,\cdots, \ \
\rho = p/q$.
 \cite{HatsugaiGap-opening,HatsugaiAnyon,Niu}

\begin{figure}[!tb]
 \includegraphics[width=7.5cm]{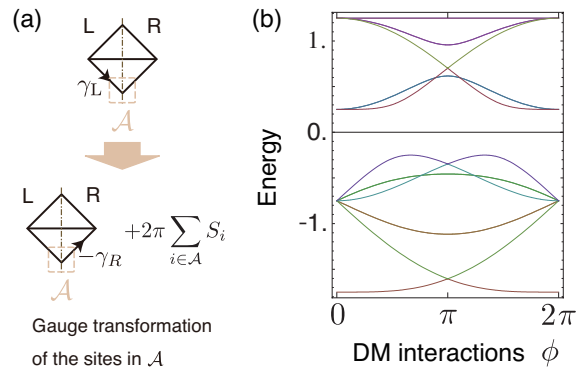}
\caption{
(a) The definition of the ${\cal A}$ in the reflection symmetric
 system, and the gauge transformation of the Berry phase. 
The system is reflection symmetric about dashed line.
(b) Spectral flow of a reflection symmetric
$S=1/2$ Heisenberg model with DM interactions $\phi$. 
The shape of lattice is the one in (a).
At $\phi=\pi$, all states are at least doubly degenerate. Some
of them are doubly degenerate for all of the twist $\phi$.
}
\label{fig:levelcross}
\label{fig:UnitLayer2}
\end{figure}

\paragraph{Reflection symmetry}
We may further apply the present argument
for the generic symmetry, such as a reflection symmetry.
Consistency between the possible Berry phases and the
reflection symmetry of the physical system requires 
some restriction.
Let us consider a reflection symmetric system consisting of two
subsystems $R$ and $L$ which are mirror images of one another.
We first choose a set of sites ${\cal A}$ where we perform 
a gauge transformation. ${\cal A}$ itself is chosen to be reflection symmetric. 
See FIG.\ref{fig:UnitLayer2}(a) for the simplest example. 
Then we define the Berry phase $\gamma_{\rm L}$ by twisting
some links on the boundary of ${\cal A}$: $\partial{\cal A}$.
Note that the self-reflection-symmetric bond, the mirror image of this bond is itself, is not twisted.
We denote this Berry phase as $\gamma_{\rm L}$, and the symmetric
partner of $\gamma_{\rm L}$ as  $\gamma_{\rm R}$.
Our gauge transformation of the Berry phase results in 
$\gamma_{\rm L}=-\gamma_{\rm R}+2\pi\sum_{j\in{\cal A}}\left(
S_j-\tilde{m}_j\right)$.
The relations $-\gamma_{\rm R}\equiv \gamma_{\rm R},\ \text{mod}\, 2\pi$ 
and $\tilde{m}_j=0$ hold
if the time reversal symmetry is also present.
In such case, we obtain that
$\gamma_{\rm L}=\gamma_{\rm R}+2\pi\sum_{{\cal A}}S_j$.
Since the reflection symmetry of the physical structure implies
$\gamma_{\rm L}\equiv \gamma_{\rm R},\ {\rm mod}\, 2\pi  $,
when the following case,
\begin{eqnarray}
\sum_{j\in{\cal A}}S_j\not\in\mathbb{Z},\nonumber
\end{eqnarray}
we predicts a level crossing during the twisting.
Our argument can be extended without any difficulty for generic reflection symmetric models including even three-dimensional ones. 
Numerical results showing the pattern of level crossings are given in
Fig.\ref{fig:levelcross}(b). Note that there is the ground state
degeneracy at $\phi=\pi$. Similarly to the translational symmetric case,
this degeneracy can also be explained by mutually anticommuting symmetry
operations $U_{\cal A}(\pi)R$ and $F$ if there is no magnetic field and $\sum_{j \in{\cal A}}S_j \not\in \mathbb{Z}$,
where $R$ denotes the reflection. 
Similar arguments can  be applied to the molecular magnets 
with Dzyaloshinsky-Moriya(DM) interactions.
We can also apply this argument 
 for the Majumdar-Ghosh model of length $4n+2$ ($n \in \mathbb{N}$) with a periodic boundary condition \cite{MG}. 
It gives a gap closing under the twist, 
which is consistent with the 
doubly degenerate  ground state at $\phi=0$.

\begin{acknowledgments}
HK was supported by the Japan Society for the Promotion of Science. 
YH was supported by Grants-in-Aid for Scientific Research
on Priority Areas from MEXT (No.18043007).
\end{acknowledgments}


\end{document}